\documentclass[12pt,a4paper,dvips]{article}
\usepackage{a4p}
\usepackage{cite,mcite}
\usepackage{graphicx}
\usepackage{physics }
\usepackage{l3_title,ifthen,Lep}
\journalname{Phys. Lett. B}
\date{November 25, 1999}
\preprint{99-176}
\Lep{2}
\newlength{\capindent}
\newlength{\capwidth}
\newlength{\figwidth}
\newcommand{\icaption}[2][!*!,!]{\hspace*{\capindent}%
  \begin{minipage}{\capwidth}
    \ifthenelse{\equal{#1}{!*!,!}}%
      {\caption{#2}}%
      {\caption[#1]{#2}}
  \end{minipage}}
\newcommand{\pho}{\phantom{0}}

\begin{document}
\begin{titlepage}

\title{Direct Observation of Longitudinally Polarised 
W$^{\pm}$ Bosons}
\author{The L3 Collaboration}
%
% The abstract
%
\begin{abstract}
\noindent
The three different helicity states of W$^{\pm}$ bosons, 
produced in the reaction 
$\mathrm{e}^{+}\mathrm{e}^{-} \rightarrow 
\mathrm{W}^{+}\mathrm{W}^{-} \rightarrow \mathrm{\ell \nu q \bar{q}}$ 
are studied using leptonic and hadronic W decays
at $\sqrt{s}$=183 \GeV{} and 189 \GeV{}. 
The W polarisation is also measured as a function of 
the scattering angle between the W$^{-}$ and the direction 
of the e$^{-}$ beam. The analysis demonstrates 
that W bosons are produced with all three helicities,  
the longitudinal and the two transverse states.
Combining the results from the two 
center--of--mass energies and with leptonic and hadronic W
decays, the fraction of longitudinally polarised W$^{\pm}$ bosons is
measured to be 0.261 $\pm$ 0.051(stat.) $\pm$ 0.016(syst.) in agreement 
with the expectation from the Standard Model.

\end{abstract}
%
% Adds "To be submitted to ..." or "Submitted to ...", if relevant
%
\submitted

\end{titlepage}

\section*{Introduction}

Previous measurements of W$^{+}$W$^{-}$ production at LEP 
have concentrated on measurements of the W mass, the W branching ratios, 
the differential and total cross sections and 
the anomalous couplings \cite{wwl31,
wwaleph1,wwopal3}. 
These measurements show, using the
differential cross sections with respect 
to the W production and decay angles, 
good agreement with theoretical calculations within  
the Standard Model~\cite{standard_model,
Veltman:1968ki}. 
This good agreement with the Standard Model indicates indirectly that 
W bosons with all three helicities are produced in the 
reaction $\mathrm{e^{+}e^{-} \rightarrow W^{+}W^{-}}$. 

The primary goal of the measurement described in this paper is
a quantitative and model independent analysis 
of all three W helicity states and in particular, the 
direct observation of longitudinally polarised W bosons.
Measurements of longitudinally polarised W bosons
have previously been reported in the reaction 
$\mathrm{e^{+}e^{-} \rightarrow W^{+}W^{-}}$~\cite{wwopal3} and
in top decays~\cite{CDF_Wlong}.
 
At center--of--mass energies close to 190 GeV and within the Standard Model,
one expects that about one quarter of all W bosons should be    
longitudinally polarised~\cite{Hagiwara87}.
Furthermore, the production of W bosons with different helicities
depends strongly on the W$^{-}$ scattering angle $\theta_{\mathrm{W}^{-}}$
with respect to the e$^{-}$ beam direction.
For example one expects for $\theta_{\mathrm{W}^{-}}$
larger than 90 degrees that almost 40\% of the events contain at least one 
longitudinally polarised W boson. In contrast, for $\theta_{\mathrm{W}^{-}}$ 
between 20 and 70 degrees, the cross section is dominated by the 
neutrino--exchange diagram and the W$^{+}$W$^{-}$ 
should be produced dominantly with transverse polarisation. 
The fractions of the W$^{\pm}$ helicity states 
should thus also be measured as a function of $\theta_{\mathrm{W}^{-}}$.

The measurement is performed with the L3 detector at LEP, 
using data samples of 55.5 pb$^{-1}$ and 176.4 pb$^{-1}$
collected at average center--of--mass energies 
of 183 \GeV{} and 189 \GeV{}, respectively.
A detailed description of the L3 detector and its performance
is given in 
reference~\cite{l3_00}.
The L3 detector response for W$^{+}$W$^{-}$
events from the KORALW~\cite{KORALW1} 
and the EEWW~\cite{EEWW} Monte Carlo programs  
is simulated with the GEANT--based L3 detector simulation 
program~\cite{xgeant}.

\section*{Analysis strategy}

The different W helicity states result in different 
angular distributions of the W decay products.
The decay angle $\theta^{*}$ in the W rest frame
between the left--handed negatively charged lepton and the W$^{-}$ 
has a $(1 \pm \cos \theta^{*})^{2}$ 
distribution for a W$^{-}$ with helicity $\mp 1$. 
The right--handed positively charged lepton has a 
$(1 \pm \cos \theta^{*})^{2}$ distribution for a W$^{+}$ 
with helicity $\pm 1$.
Longitudinally polarised W bosons (helicity $0$)  
result in a symmetric distribution of the decay products,
proportional to $\sin^{2} \theta^{*}$.
To simplify the description of the 
helicity fractions, we refer in the following text only 
to the fractions $f_{-}$, $f_{0}$ and $f_{+}$ of the W$^{-}$ helicities, which 
includes the corresponding  
W$^{+}$ states with $f_{+}$, $f_{0}$ and $f_{-}$, respectively.
 
In order to study the W polarisation, we use events of the type 
$\mathrm{e^{+}e^{-} \rightarrow W^{+}W^{-} \rightarrow 
\ell^{\pm} \nu q \bar{q}}$
with $\mathrm{\ell}^{\pm}$ being either e$^{\pm}$ or $\mu^{\pm}$. The neutrino 
four--momentum vector is reconstructed from the 
total missing momentum vector of the event.
These event samples are essentially background free and allow a measurement 
with good accuracy of the W$^{\pm}$ momentum vector, the W charge and the 
decay angle $\theta^{*}$ in the W rest frame.

In contrast to leptonic W decays, where 
the decay angle $\theta^{*}_{\ell}$ of the $\ell^{\pm}$
is well defined, the corresponding $\theta^{*}_{q}$ for quarks  
in W decays has to be calculated from the hadronic decay products. 
To approximate the quark decay angle in the W rest frame, we proceed 
in the following way. First,  
all particles besides the charged lepton and the missing neutrino 
in the event are associated with the hadronic decay of the W. 
We then calculate their associated four--vectors
in the rest frame of the W and determine the corresponding   
thrust axis in this rest frame. The angle $\theta^{*}_{\mathrm{Thrust}}$
of this thrust axis
with respect to the W momentum vector in the laboratory frame is 
used to describe the quark decay angle $\theta^{*}_{q}$ in the W 
rest frame. 

After correcting for efficiencies, the contributions from 
different W polarisation states are  
obtained from a fit to the $\cos \theta^{*}$ distributions. 
For the leptonic W decays the fractions $f_{-}$, $f_{+}$ and 
$f_{0}$ of the three W helicity states are 
obtained from:  

\begin{eqnarray}
\frac{1}{N}\frac{dN}{d\cos\theta^{*}} = f_{-}  
\frac{3}{8}~(1+\cos\theta^{*})^{2} + 
f_{+} \frac{3}{8}~(1-\cos\theta^{*})^{2} + f_{0} 
\frac{3}{4} \sin^{2} \theta^{*}.
\end{eqnarray}

For hadronic W decays, without quark charge identification,
one measures only the absolute value of the W hadronic decay angle 
$|\cos \theta^{*}|$. However, 
this distribution can still be used  
to measure the fractions for the sum of the two transverse helicity 
states $f_{\pm}= f_{-} + f_{+}$ and  
$f_{0}$ using: 

\begin{eqnarray}
\frac{1}{N}\frac{dN}{d|\cos\theta^{*}|} = f_{\pm}  
\frac{3}{4} (1+\cos^{2}\theta^{*}) 
+ f_{0}  \frac{3}{2} \sin^{2} \theta^{*}.
\end{eqnarray}

The predictions for the compositions of W helicity states as 
a function of the W$^{-}$ scattering angle $\theta_{\mathrm{W}^{-}}$, 
following the formalism of Hagiwara {\it et al.}~\cite{Hagiwara87} 
and its implementation in the KORALW Monte Carlo 
program~\cite{KORALW1}, are used as the Standard Model 
prediction for our analysis. 
The helicity composition of the total W sample 
is extracted from a fit to the distribution of
the simulated decay angles.
From a fit to a KORALW Monte Carlo event sample at $\sqrt{s}$=189 \GeV{},
with a size 100 times larger than the data sample, 
the Standard Model predictions for 
inclusive W helicity fractions $f_{-}$, $f_{+}$ and $f_{0}$ 
are obtained to be 56.3\%, 18.0\% and 25.7\%, respectively.
The statistical errors are smaller than 0.5\%.

Within the statistical errors, the same fractions are found from 
a WW event sample generated with the EEWW 
Monte Carlo program~\cite{EEWW} which uses the zero  
total W width approximation and assigns the W helicities 
on an event by event basis.
The W helicity fractions obtained from the fit to the 
decay angle distributions agree, 
within statistical errors smaller than 0.9\%, with the 
generated W helicity fractions. This shows that the Born level 
formul\ae{} (1) and (2) are applicable after radiative corrections. 

\section*{Selection of {\boldmath$ 
{\mathrm {W^{+}W^{-} \rightarrow e (\mu) \nu q\bar{q}}}$} events}

The selection of $\mathrm{W^{+}W^{-} \rightarrow e (\mu) \nu q\bar{q}}$ events
is similar to the  selections described in our previous publications 
on WW final states~\cite{wwl31,*wwl32}. However, in order to assure 
well measured W production and decay angles, 
more restrictive criteria are used.
Charged leptons are identified using their characteristic
signatures.  Electrons are identified as isolated energy depositions
in the electromagnetic calorimeter with electromagnetic shower
shape which are matched in azimuth to a track reconstructed in the
central tracking chamber. The energy and direction of electrons are
measured using the electromagnetic calorimeter, 
while the charge is obtained from the associated track. 
Muons are identified and measured with tracks
reconstructed in the muon chambers which point back to the interaction
vertex. All other energy depositions in the calorimeters are assumed to 
originate from the hadronic W decay. The neutrino 
momentum vector is set equal to the total missing momentum vector of the 
event. In addition the following criteria are used for the selection of 
$\mathrm{W^{+}W^{-} \rightarrow e (\mu) \nu q\bar{q}}$ events:
\begin{itemize}
\item 
The reconstructed momentum 
should be greater than 20 \GeV{} for electrons and 15 \GeV{} for muons.
\item
The momentum of the neutrino should be greater than 10 \GeV{} 
and its polar angle, $\theta_{\nu}$, 
has to satisfy $|\cos \theta_{\nu}| < 0.95$.
\item 
The invariant mass of the $\ell \nu$ system should be 
greater than 60 \GeV{}.

\item
The invariant mass of the hadronic system should be 
between 50 and 110 \GeV{}.
\end{itemize}

Using these criteria, 81 and 288 events of the type 
$\mathrm{W^{+}W^{-} \rightarrow e \nu q\bar{q}}$
are selected at center--of--mass energies of 183 \GeV{} and 
189 \GeV{}, respectively. The corresponding event numbers for 
$\mathrm{\mu \nu q\bar{q}}$ 
are 67 and 262 events. 
Adding the electron and muon event samples together,
we find 68 and 280 $\ell^{-}$ events and 80 and 270 $\ell^{+}$ events, 
respectively, in the 183 \GeV{} and 189 \GeV{} data samples.   
These samples have a purity of 96\%, where 
the background from $\mathrm{W^{+}W^{-} \rightarrow \tau \nu q\bar{q}}$
with leptonic $\tau$ decays 
and the background from $\mathrm{e^{+}e^{-}} \rightarrow$ hadrons
contribute each about 2\%. 

\begin{figure}[htbp]
  \begin{center}
    \includegraphics[width=\figwidth]{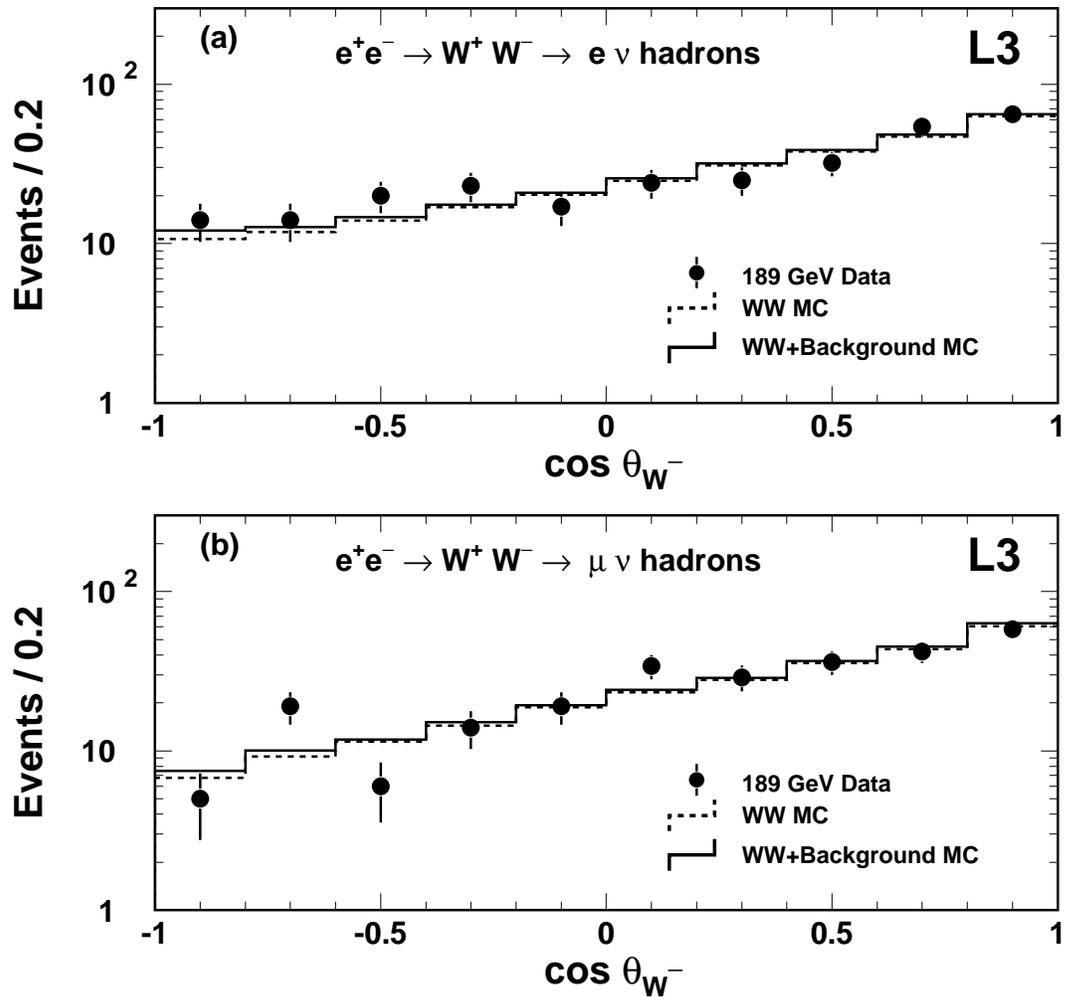}
  \end{center}
  \caption{The $\cos \theta_{\mathrm{W}^{-}}$ distribution for (a)
$\mathrm{W^{+}W^{-} \rightarrow e \nu q\bar{q}}$ and (b) 
$\mathrm{W^{+}W^{-} \rightarrow \mu \nu q\bar{q}}$ events 
from the $\sqrt{s}=189$ \GeV{} data (points) 
and the KORALW Monte Carlo prediction (histogram).} 
\label{fig:wwpol1}
\end{figure}

The measured $\cos \theta_{\mathrm{W}^{-}}$ distribution
is found to be in good agreement with the MC expectations, 
as shown in Figure~\ref{fig:wwpol1} for events with electrons and 
with muons for the 189 \GeV{} data sample.
About 5\% of the accepted events with 
electron candidates have a wrongly assigned charge. Charge confusion is
insignificant for events with muons. 
The charge confusion depends on the reconstructed W$^{-}$ 
scattering angle and is largest for W bosons with small scattering 
angle with respect to the beam direction. This results in a 
small misassignment between W bosons with helicity +1 and --1 but has 
negligible effects for the fraction of longitudinally 
polarised W bosons, which is essentially independent of 
the charge assignment.

\section*{Analysis of the W helicity states}
After subtracting the backgrounds from the data, 
the fractions of the W helicity states are measured 
from the distributions $dN/d\cos \theta^{*}_{\ell^{\pm}}$ 
and $dN/ d|\cos \theta^{*}_{\mathrm{Thrust}}|$  for the leptonic 
and hadronic W decay angle and 
as a function of the scattering angle $\theta_{\mathrm{W}^{-}}$.

To extract the W helicity fractions, 
the observed distributions   
are corrected for the selection 
efficiencies which are obtained as a function of $\cos \theta^{*}$.
To take into account possible deviations between 
the helicity fractions in the data and Monte Carlo as a function 
of $\cos \theta_{\mathrm{W}^{-}}$, the 
data are corrected differentially using 9 bins 
of the $\cos \theta_{\mathrm{W}^{-}}$ scattering angle. 
For each $\cos \theta_{\mathrm{W}^{-}}$ bin, the efficiency 
is obtained as a function of $\cos \theta^{*}$ using the 
ratio of the reconstructed and the generated 
$\cos \theta^{*}$ distributions for the leptonic and 
hadronic W decays. The measured $\cos \theta^{*}$ distributions 
for the corresponding $\cos \theta_{\mathrm{W}^{-}}$ bins in the 
data are corrected and combined. 

The efficiency corrections are obtained from large samples of 
fully simulated 
KORALW Monte Carlo events.
Using these Monte Carlo events we have studied
the accuracy with which we reconstruct the $\theta^{*}$ decay angles. 
The study shows that $\theta^{*}$ is reconstructed with a standard 
deviation of 9.2 degrees and a small shift of $-$3.2
degrees for the leptonic W decays. For hadronic W decays 
one finds that $\theta^{*}$ is reconstructed with a standard 
deviation of 12.0 degrees and a shift of $+$3.3 degrees.

The bias and sensitivity loss due to the efficiency corrections and 
the $\theta^{*}$ resolution has been determined 
with fully simulated and reconstructed Monte Carlo events where
the generated W helicity fractions have been varied over a large range.
This was done both with the EEWW Monte Carlo program, where
the generated W helicities are known on an event 
by event basis and with the KORALW Monte Carlo using 
a weighting method to assign the W helicities on a 
statistical basis, ignoring W spin correlations.

Averaging both Monte Carlo estimates   
one finds that leptonic W decays with 100\% helicity $-1$ states would 
be measured to consist of 94\% of helicity $-1$ and 6\% helicity $0$ 
states while a W sample with 100\% helicity $+1$
would be reconstructed to consist of more than 99\% of 
helicity $+1$ states.
Similar numbers are found if one starts with 100\% helicity $0$, which 
would be measured with 92\% helicity $0$ , 3\% helicity $-1$ 
and 5\% helicity $+1$.
The corresponding numbers for hadronic W decays are that 94\% of W bosons 
with helicity $\pm 1$ and 85\% of W bosons with helicity $0$ are correctly 
reconstructed. The study has been repeated 
as a function of $\cos \theta_{\mathrm{W}^{-}}$ and 
within the statistical errors the 
results are the same as the ones from the total W sample.
To obtain a correction function for the bias and the 
efficiency loss, the fraction $f_{0}$ 
has been varied between 0 and 100\%. 
A linear relation between the generated and the fitted W 
helicity fractions is found. 

\subsection*{Results and systematics}

\begin{figure}[htbp]
  \begin{center}
    \includegraphics[width=\figwidth]{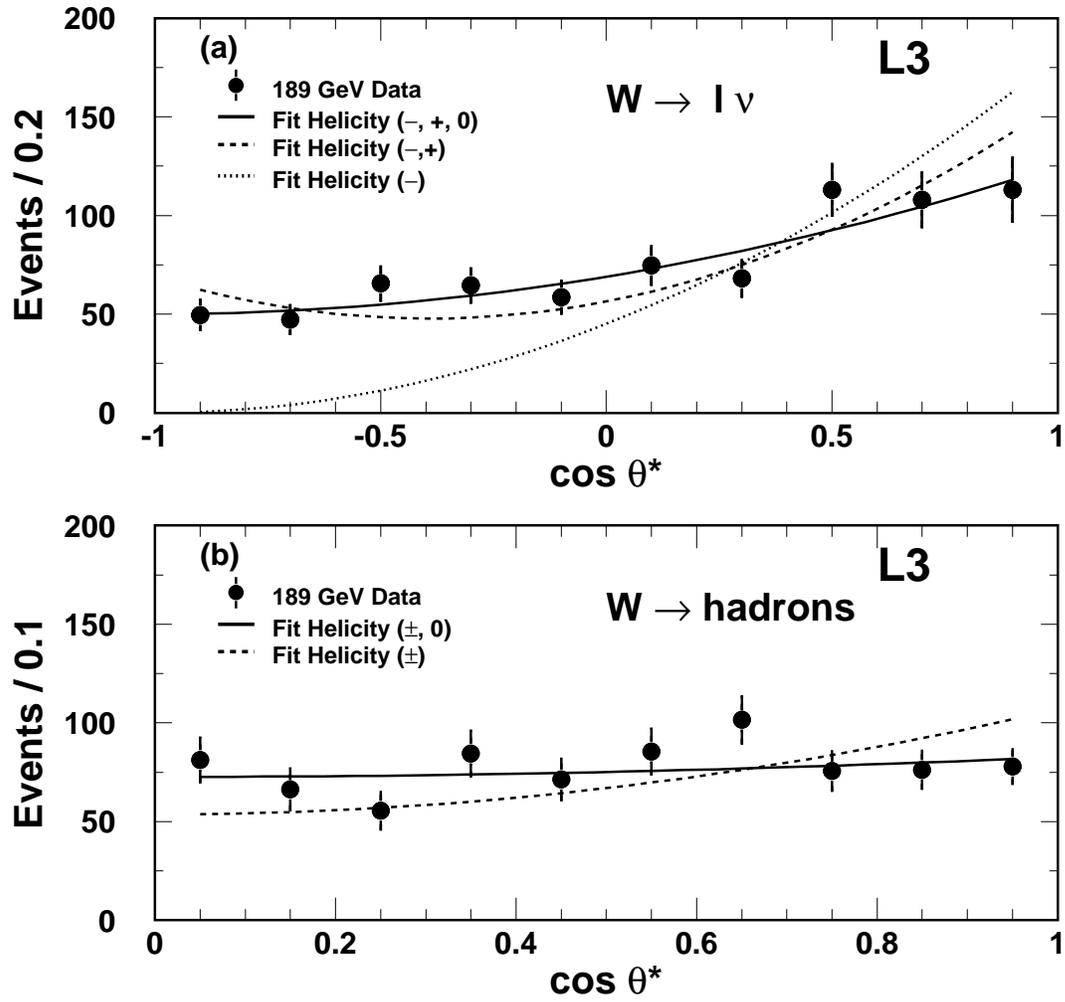}
  \end{center}
  \caption{Efficiency-- and background--corrected
$\cos \theta^{*}$ distributions for (a) leptonic W decays
and (b) for hadronic W decays at $\sqrt{s}=189$ \GeV{}. The
fit results for the different W helicity hypotheses are also shown.
} 
\label{fig:wwpol2}
\end{figure}

These efficiency corrected   
$\cos \theta^{*}$ distributions are used to extract 
the W$^{\pm}$ helicity fractions.
The results of the binned $\chi^{2}$ fits to these distributions for 
leptonic and hadronic W decays from 
the $\sqrt{s}$=189 \GeV{} data 
are shown in Figure~\ref{fig:wwpol2}. No constraint on the total 
cross section is applied and one finds that 
the data are well described 
only if all three W helicity states are used in the fit.  
Fits which include only 
$-1$ and $\pm 1$ helicities, as also shown 
in Figure 2, fail to describe the data.
For leptonic W decays one finds that 
the $\chi^{2}$ increases from 7.1 for seven degrees of freedom 
if all three W helicity states
are included to 17.8 for eight degrees of freedom
if only helicity $-1$ and $+1$ are used to describe the data.
For hadronic W decays the 
$\chi^{2}$ increases from 9.8 for eight degrees of freedom 
if all three W helicity states are included to 26.4 for 
nine degrees of freedom
if only helicity $\pm 1$ are used to describe the data. 

The fraction of longitudinally polarised W bosons in the 
$\sqrt{s}$=189 \GeV{} data is measured to be 
0.220$\pm$0.077 for the leptonic decays and 0.285$\pm$0.084 for hadronic 
decays. The fractions for the different W helicity states,
together with the Standard Model Monte Carlo expectations, are given 
in Table 1 for the $\sqrt{s}$=189 \GeV{} and $\sqrt{s}$=183 \GeV{} data.
The observed fractions of longitudinally polarised W bosons
measured with leptonic and hadronic W decays agree with each other 
and with the Standard Model 
expectation of 0.26 and differ from zero by several standard 
deviations.

Systematic studies have been performed to  
verify the stability of the fit results with respect to the 
fraction of longitudinally polarised W bosons. We have investigated 
(1) uncertainties due to backgrounds, 
(2) efficiencies and selection criteria, 
(3) the hadron energy response functions of the 
electromagnetic and hadronic calorimeters,
(4) the difference between the differential
and overall efficiency corrections and 
(5) a method where the fraction $f_{0}$
has been obtained directly from a fit to the measured $\cos \theta^{*}$
distributions using the Monte Carlo shape from the different 
W helicity states after the reconstruction.  

The analysis has been repeated assuming 
large relative background uncertainies of $\pm$ 50\%
from either the hadronic background or from misidentified 
W$ \rightarrow \tau \nu$ decays.  
Using these modifications the measured fractions of longitudinally polarised
W bosons is found to vary by at most 0.012 for leptonic W decays and 
by 0.004 for the hadronic W decays. The hadron energy measurement 
is obtained from a combination of the energy deposited in the 
electromagnetic and hadron calorimeter multiplied by
calibration constants which take the average calorimeter e$^{\pm}$/hadron 
response function into account. 
These calibration constants have been varied over a wide range
while demanding that  
the average of the reconstructed masses 
for leptonic and hadronic W decays agree
within better than $\pm$ 3 \GeV{} with an average W mass of 
80.4 \GeV{}. Since the neutrino momentum vector is reconstructed 
from the observed missing momentum vector, correlations exist  
between the reconstructed decay angles 
in the hadronic W system and the corresponding leptonic 
W system. For example, a particular choice of the 
energy calibration constants reduces the fraction of longitudinally 
polarised W bosons by 0.024 as seen with the leptonic W decays 
but increases the corresponding fraction 
for the hadronic decays by 0.015. 

Similar variations in the fraction of 
longitudinally polarised W bosons have been seen with the other 
systematic studies, as summarised in Table 2.  
Assuming that the variations given in Table 
2 are all due to systematics and adding them in quadrature, 
a systematic error of $\pm$0.034, $\pm$0.024 and $\pm$0.016 
is assigned to the fraction of longitudinally 
polarised W bosons measured with leptonic, hadronic  
decays and for the combined measurement, respectively. 

Combining the results from the $\sqrt{s} =$ 183 \GeV{} and 189 \GeV{}, 
ignoring the slight energy dependence of the W helicity fractions
expected from the Standard Model, the fraction of longitudinally 
polarised W bosons is measured to be 

\begin{eqnarray}
f_{0} = 0.261 \pm 0.051(stat.) \pm 0.016(syst.) \nonumber
\end{eqnarray}

\noindent
and agrees with expectation from the Standard Model of 0.26.

\begin{figure}[htbp]
  \begin{center}
    \includegraphics[width=\figwidth]{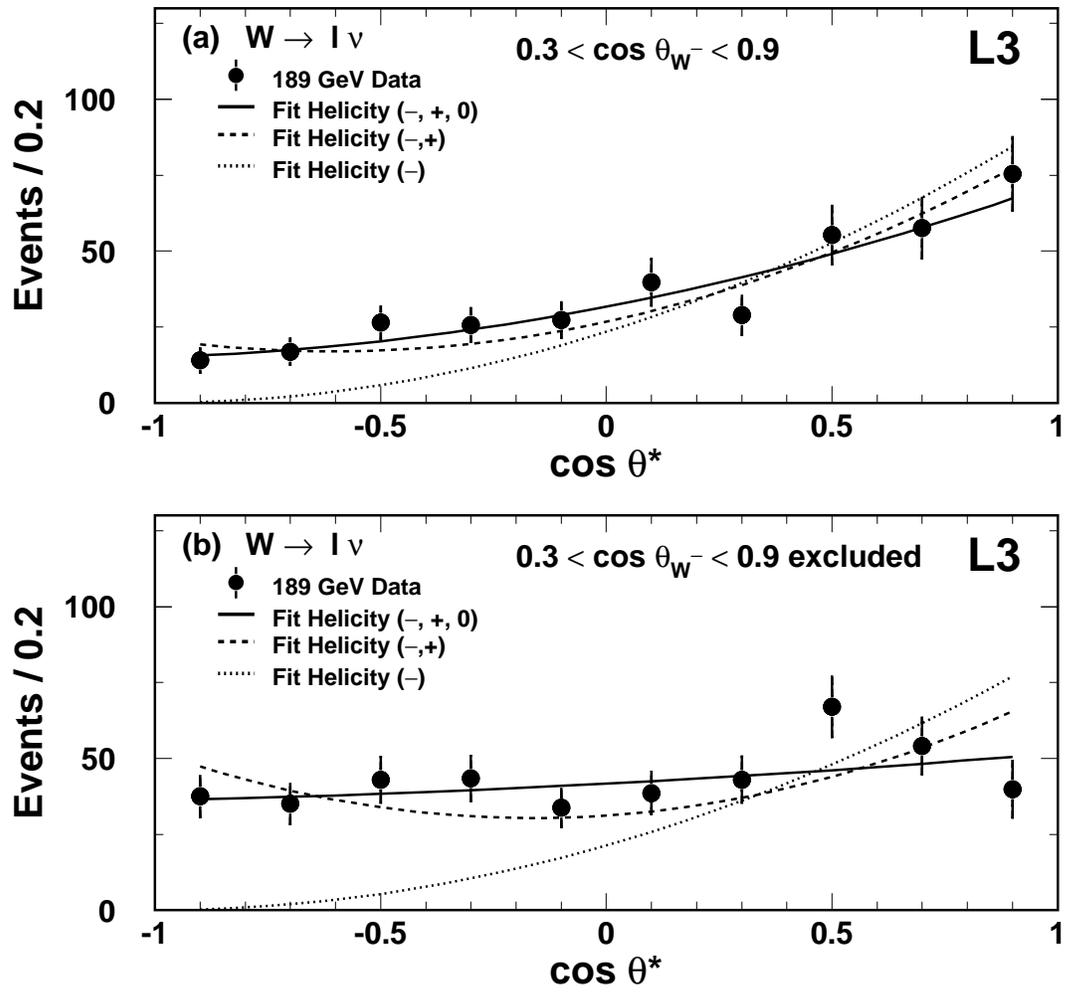}
  \end{center}
  \caption{Corrected $\cos \theta^{*}$ distribution 
from leptonic W decays for (a) enriched and (b) depleted
transverse W polarisation regions together with the
fit results.
For (a) the required $\theta_{\mathrm{W}^{-}}$ must satisfy
$0.3 < \cos \theta_{\mathrm{W}^{-}}< 0.9$, while for (b) it has to be 
$\cos \theta_{\mathrm{W}^{-}}< 0.3$ or 
$ 0.9 < \cos \theta_{\mathrm{W}^{-}}$.
} 
\label{fig:wwpol3}
\end{figure}

\begin{figure}[htbp]
  \begin{center}
    \includegraphics[width=\figwidth]{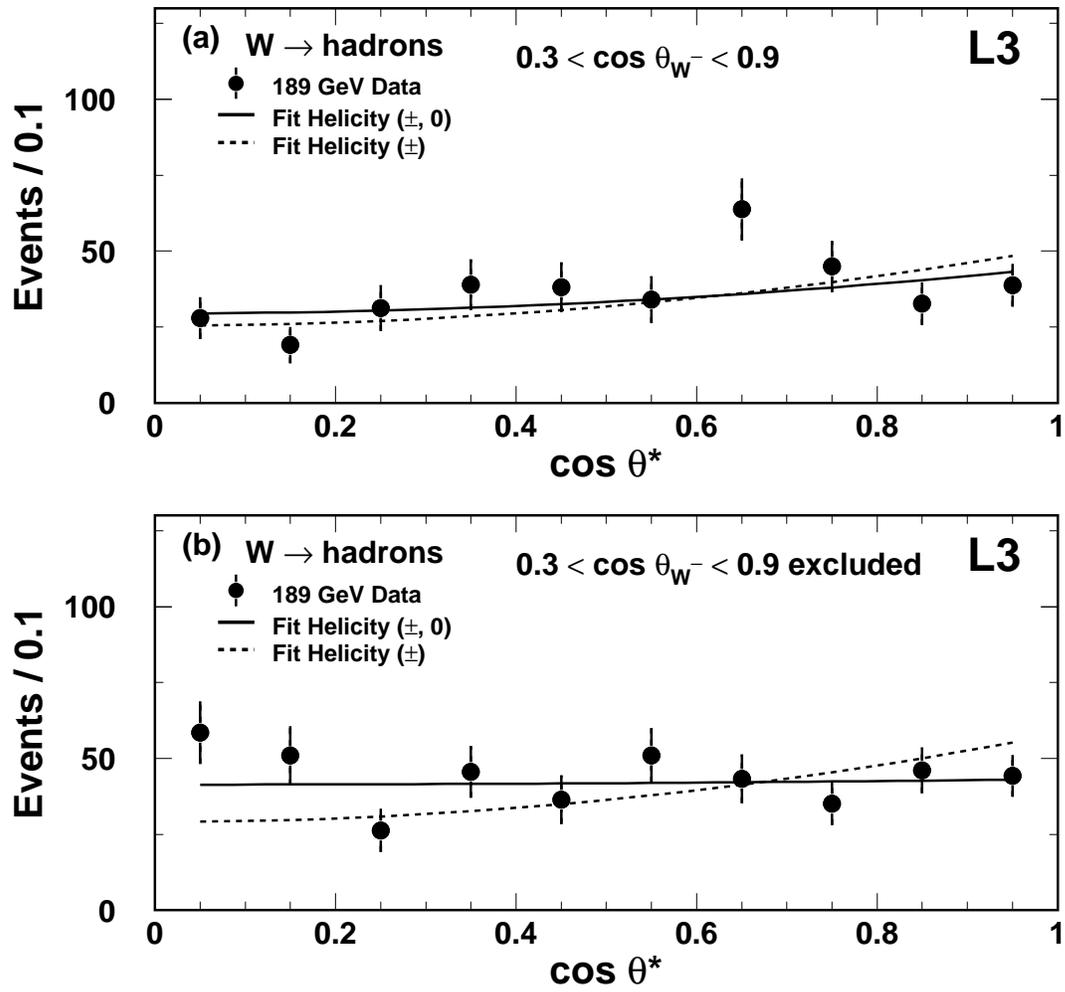}
  \end{center}
  \caption{Same as Figure 3 except that in this case 
$|\cos \theta^{*}|$ is shown for hadronic W decays.
} 
\label{fig:wwpol4}
\end{figure}

As mentioned in the introduction, it is  
interesting to measure the W helicity fractions as a function of the 
W$^{-}$ scattering angle $\theta_{\mathrm{W}^{-}}$.
Thus the fits are repeated for different ranges of 
$\cos \theta_{\mathrm{W}^{-}}$.
The $\cos \theta_{\mathrm{W}^{-}}$ ranges are selected such that 
the contributions from the transversely polarised W bosons
should be either suppressed or enhanced 
as shown in Figures~\ref{fig:wwpol3} and~\ref{fig:wwpol4}.

\begin{figure}[htbp]
  \begin{center}
    \includegraphics[width=\figwidth]{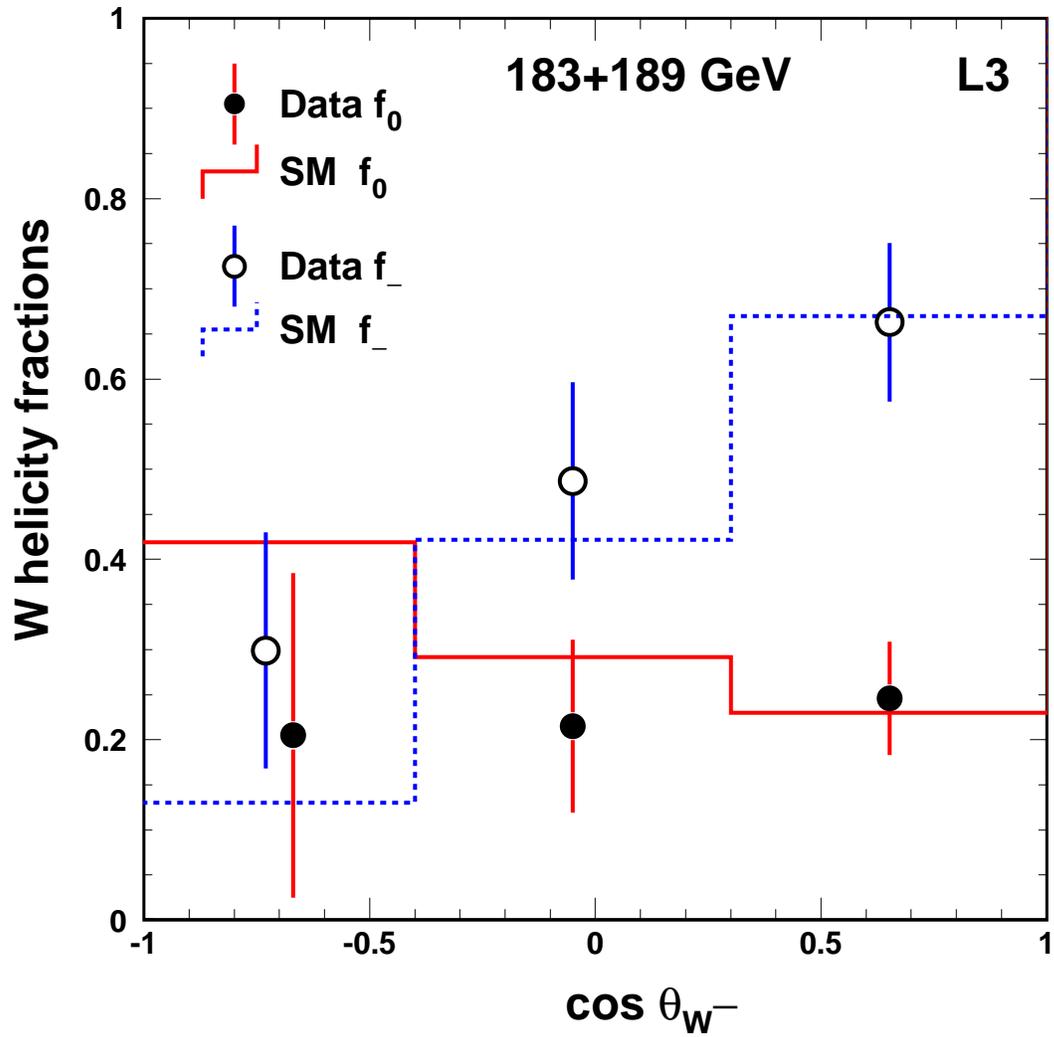}
  \end{center}
  \caption{W helicity fractions $f_{0}$ and $f_{-}$ 
for three different bins of $\cos \theta_{\mathrm{W}^{-}}$
in the combined data sample and in the KORALW Monte Carlo.  
} 
\label{fig:wwpol5}
\end{figure}

To obtain quantitative numbers for the W helicity fractions 
as a function of $\cos \theta_{\mathrm{W}^{-}}$ 
the data from the two different center--of--mass energies 
are combined and the helicity fractions are measured 
for three bins of $\cos \theta_{\mathrm{W}^{-}}$.  
The bins are chosen such that large variations of the different 
helicity fractions are expected \cite{Hagiwara87} 
yet keeping a sufficient statistical significance.
The results, given in Table 3, agree with the Standard Model 
expectations and demonstrate that the fraction of 
W bosons with helicity $-1$ depends on the W scattering 
angle as shown in Figure~\ref{fig:wwpol5}.

In summary, all three W boson helicity states, 
the two transverse as well as the 
longitudinal ones are observed with fractions in agreement with 
Standard Model expectations.
The production of longitudinally polarised W bosons 
is thus directly observed with a significance of five standard deviations.
      
%
%%%%%%%%%%%%%%%%%%%%%%%%%%%%%%%%%%%%%%%%%%%%%%%%%%%%%%%%%%%%%%%%%%%%%%%%%%%%%%%
% Acknowledgements
%%%%%%%%%%%%%%%%%%%%%%%%%%%%%%%%%%%%%%%%%%%%%%%%%%%%%%%%%%%%%%%%%%%%%%%%%%%%%%%
%
\section*{Acknowledgments}

We would like to thank F. Jegerlehner, Z. Kunszt, 
Z. W\c{a}s and D. Zeppenfeld for interesting discussions 
about the physics of WW production. 

We also wish to 
express our gratitude to the CERN accelerator divisions for
the excellent performance of the LEP machine. 
We acknowledge the contributions of the engineers 
and technicians who have participated in the construction 
and maintenance of this experiment.  
\newpage
%
%%%%%%%%%%%%%%%%%%%%%%%%%%%%%%%%%%%%%%%%%%%%%%%%%%%%%%%%%%%%%%%%%%%%%%%%%%%%%%%
% Bibliography
%%%%%%%%%%%%%%%%%%%%%%%%%%%%%%%%%%%%%%%%%%%%%%%%%%%%%%%%%%%%%%%%%%%%%%%%%%%%%%
%
% Style file to use with mcite.
% Use l3style with just cite.
\bibliographystyle{l3stylem}

\newpage
\typeout{   }     
\typeout{Using author list for paper 195 -?}
\typeout{$Modified: Tue Nov 23 09:52:18 1999 by clare $}
\typeout{!!!!  This should only be used with document option a4p!!!!}
\typeout{   }
%
%
%
%  L A T E X  version!!
%
%
% Make sure that the Lep package has been used!
%\input{Lep.sty}%
%
%\ifx\LepCalled\undefined%
%\typeout{     }%
%\typeout{!!!!!!!!!!!!!!!!!!!!!!!!!!!!!!!!!!!!!!!!!!!!!!!!!!!!!!!!!!!}%
%\typeout{Yikes.  You haven't used the Lep package!}%
%\typeout{Please put \protect\usepackage\protect{Lep\protect} in your preamble,
%         followed by}%
%\typeout{\protect\Lep\protect{1\protect} or \protect\Lep\protect{2\protect}}%
%\typeout{     }%
%\typeout{For now you will get a Lep phase 2 authorlist (may not be right!).}%
%\typeout{!!!!!!!!!!!!!!!!!!!!!!!!!!!!!!!!!!!!!!!!!!!!!!!!!!!!!!!!!!!}%
%\typeout{     }%
%\Lep{2}\fi%

\newcount\tutecount  \tutecount=0
\def\tutenum#1{\global\advance\tutecount by 1 \xdef#1{\the\tutecount}}
\def\tute#1{$^{#1}$}
\tutenum\aachen            % 1
\tutenum\nikhef            % 2
\tutenum\mich              % 3
\tutenum\lapp              % 4
\tutenum\basel             % 5
\tutenum\lsu               % 6
\tutenum\beijing           % 7
\tutenum\berlin            % 8
\tutenum\bologna           % 9 
\tutenum\tata              % 10
\tutenum\ne                % 11
\tutenum\bucharest         % 12
\tutenum\budapest          % 13
\tutenum\mit               % 14 
\tutenum\debrecen          % 15
\tutenum\florence          % 16
\tutenum\cern              % 17 
\tutenum\wl                % 18 
\tutenum\geneva            % 19
\tutenum\hefei             % 20
\tutenum\seft              % 21
\tutenum\lausanne          % 22
\tutenum\lecce             % 23
\tutenum\lyon              % 24
\tutenum\madrid            % 25
\tutenum\milan             % 26
\tutenum\moscow            % 27
\tutenum\naples            % 27
\tutenum\cyprus            % 29
\tutenum\nymegen           % 30
\tutenum\caltech           % 31
\tutenum\perugia           % 32
\tutenum\cmu               % 33
\tutenum\prince            % 34
\tutenum\rome              % 35
\tutenum\peters            % 36
\tutenum\salerno           % 37
\tutenum\ucsd              % 38
\tutenum\santiago          % 39
\tutenum\sofia             % 40
\tutenum\korea             % 41
\tutenum\alabama           % 42
\tutenum\utrecht           % 43
\tutenum\purdue            % 44
\tutenum\psinst            % 45
\tutenum\zeuthen           % 46
\tutenum\eth               % 47
\tutenum\hamburg           % 48
\tutenum\taiwan            % 49
\tutenum\tsinghua          % 50
{
\parskip=0pt
\noindent
{\bf The L3 Collaboration:}
\ifx\selectfont\undefined%  old style font selection
 \baselineskip=10.8pt
 \baselineskip\baselinestretch\baselineskip
 \normalbaselineskip\baselineskip
 \ixpt
\else%                      new style font selection
 \fontsize{9}{10.8pt}\selectfont
\fi
\medskip
\tolerance=10000
\hbadness=5000
\raggedright
\hsize=162truemm\hoffset=0mm
\def\r{\rlap,}
\noindent

M.Acciarri\r\tute\milan\
P.Achard\r\tute\geneva\ 
O.Adriani\r\tute{\florence}\ 
M.Aguilar-Benitez\r\tute\madrid\ 
J.Alcaraz\r\tute\madrid\ 
G.Alemanni\r\tute\lausanne\
J.Allaby\r\tute\cern\
A.Aloisio\r\tute\naples\ 
M.G.Alviggi\r\tute\naples\
G.Ambrosi\r\tute\geneva\
% ...
\newpage
%\rule{\textwidth}{0.4pt}
\begin{list}{A}{\itemsep=0pt plus 0pt minus 0pt\parsep=0pt plus 0pt minus 0pt
                \topsep=0pt plus 0pt minus 0pt}
\item[\aachen]
 I. Physikalisches Institut, RWTH, D-52056 Aachen, FRG$^{\S}$\\
 III. Physikalisches Institut, RWTH, D-52056 Aachen, FRG$^{\S}$
\item[\nikhef] National Institute for High Energy Physics, NIKHEF, 
     and University of Amsterdam, NL-1009 DB Amsterdam, The Netherlands
\item[\mich] University of Michigan, Ann Arbor, MI 48109, USA
\item[\lapp] Laboratoire d'Annecy-le-Vieux de Physique des Particules, 
     LAPP,IN2P3-CNRS, BP 110, F-74941 Annecy-le-Vieux CEDEX, France
\item[\basel] Institute of Physics, University of Basel, CH-4056 Basel,
     Switzerland
\item[\lsu] Louisiana State University, Baton Rouge, LA 70803, USA
\item[\beijing] Institute of High Energy Physics, IHEP, 
  100039 Beijing, China$^{\triangle}$ 
\item[\berlin] Humboldt University, D-10099 Berlin, FRG$^{\S}$
\item[\bologna] University of Bologna and INFN-Sezione di Bologna, 
     I-40126 Bologna, Italy
\item[\tata] Tata Institute of Fundamental Research, Bombay 400 005, India
\item[\ne] Northeastern University, Boston, MA 02115, USA
\item[\bucharest] Institute of Atomic Physics and University of Bucharest,
     R-76900 Bucharest, Romania
\item[\budapest] Central Research Institute for Physics of the 
     Hungarian Academy of Sciences, H-1525 Budapest 114, Hungary$^{\ddag}$
\item[\mit] Massachusetts Institute of Technology, Cambridge, MA 02139, USA
\item[\debrecen] KLTE-ATOMKI, H-4010 Debrecen, Hungary$^\P$
\item[\florence] INFN Sezione di Firenze and University of Florence, 
     I-50125 Florence, Italy
\item[\cern] European Laboratory for Particle Physics, CERN, 
     CH-1211 Geneva 23, Switzerland
\item[\wl] World Laboratory, FBLJA  Project, CH-1211 Geneva 23, Switzerland
\item[\geneva] University of Geneva, CH-1211 Geneva 4, Switzerland
\item[\hefei] Chinese University of Science and Technology, USTC,
      Hefei, Anhui 230 029, China$^{\triangle}$
\item[\seft] SEFT, Research Institute for High Energy Physics, P.O. Box 9,
      SF-00014 Helsinki, Finland
\item[\lausanne] University of Lausanne, CH-1015 Lausanne, Switzerland
\item[\lecce] INFN-Sezione di Lecce and Universit\'a Degli Studi di Lecce,
     I-73100 Lecce, Italy
\item[\lyon] Institut de Physique Nucl\'eaire de Lyon, 
     IN2P3-CNRS,Universit\'e Claude Bernard, 
     F-69622 Villeurbanne, France
\item[\madrid] Centro de Investigaciones Energ{\'e}ticas, 
     Medioambientales y Tecnolog{\'\i}cas, CIEMAT, E-28040 Madrid,
     Spain${\flat}$ 
\item[\milan] INFN-Sezione di Milano, I-20133 Milan, Italy
\item[\moscow] Institute of Theoretical and Experimental Physics, ITEP, 
     Moscow, Russia
\item[\naples] INFN-Sezione di Napoli and University of Naples, 
     I-80125 Naples, Italy
\item[\cyprus] Department of Natural Sciences, University of Cyprus,
     Nicosia, Cyprus
\item[\nymegen] University of Nijmegen and NIKHEF, 
     NL-6525 ED Nijmegen, The Netherlands
\item[\caltech] California Institute of Technology, Pasadena, CA 91125, USA
\item[\perugia] INFN-Sezione di Perugia and Universit\'a Degli 
     Studi di Perugia, I-06100 Perugia, Italy   
\item[\cmu] Carnegie Mellon University, Pittsburgh, PA 15213, USA
\item[\prince] Princeton University, Princeton, NJ 08544, USA
\item[\rome] INFN-Sezione di Roma and University of Rome, ``La Sapienza",
     I-00185 Rome, Italy
\item[\peters] Nuclear Physics Institute, St. Petersburg, Russia
\item[\salerno] University and INFN, Salerno, I-84100 Salerno, Italy
\item[\ucsd] University of California, San Diego, CA 92093, USA
\item[\santiago] Dept. de Fisica de Particulas Elementales, Univ. de Santiago,
     E-15706 Santiago de Compostela, Spain
\item[\sofia] Bulgarian Academy of Sciences, Central Lab.~of 
     Mechatronics and Instrumentation, BU-1113 Sofia, Bulgaria
\item[\korea] Center for High Energy Physics, Adv.~Inst.~of Sciences
     and Technology, 305-701 Taejon,~Republic~of~{Korea}
\item[\alabama] University of Alabama, Tuscaloosa, AL 35486, USA
\item[\utrecht] Utrecht University and NIKHEF, NL-3584 CB Utrecht, 
     The Netherlands
\item[\purdue] Purdue University, West Lafayette, IN 47907, USA
\item[\psinst] Paul Scherrer Institut, PSI, CH-5232 Villigen, Switzerland
\item[\zeuthen] DESY, D-15738 Zeuthen, 
     FRG
\item[\eth] Eidgen\"ossische Technische Hochschule, ETH Z\"urich,
     CH-8093 Z\"urich, Switzerland
\item[\hamburg] University of Hamburg, D-22761 Hamburg, FRG
\item[\taiwan] National Central University, Chung-Li, Taiwan, China
\item[\tsinghua] Department of Physics, National Tsing Hua University,
      Taiwan, China
\item[\S]  Supported by the German Bundesministerium 
        f\"ur Bildung, Wissenschaft, Forschung und Technologie
\item[\ddag] Supported by the Hungarian OTKA fund under contract
numbers T019181, F023259 and T024011.
\item[\P] Also supported by the Hungarian OTKA fund under contract
  numbers T22238 and T026178.
\item[$\flat$] Supported also by the Comisi\'on Interministerial de Ciencia y 
        Tecnolog{\'\i}a.
\item[$\sharp$] Also supported by CONICET and Universidad Nacional de La Plata,
        CC 67, 1900 La Plata, Argentina.
\item[$\diamondsuit$] Also supported by Panjab University, Chandigarh-160014, 
        India.
\item[$\triangle$] Supported by the National Natural Science
  Foundation of China.
\item[\dag] Deceased.
\end{list}
}
\vfill

%%% Local Variables: 
%%% mode: latex
%%% TeX-master: t
%%% End:

%%% Local Variables: 
%%% mode: latex
%%% TeX-master: t
%%% TeX-master: t
%%% TeX-master: t
%%% End: 

%%% Local Variables: 
%%% mode: latex
%%% TeX-master: t
%%% End: 

%%% Local Variables: 
%%% mode: latex
%%% TeX-master: t
%%% End: 

%%% Local Variables: 
%%% mode: latex
%%% TeX-master: t
%%% End: 

\newpage

%%%%%%%%%%%%%%%%%%%%%%%%%%%%%%%%%%%%%%%%%%%%%%%%%%%%%%%%%%%%%%%%%%%%%%%%%%%%%%
% \subsection*{Tables}

\begin{table}[htbp]
  \begin{center}
    \begin{tabular}{|c|c|c|c||c|c|}\hline
        \multicolumn{6}{|c|}{$\sqrt{s}=189$ \GeV{}} \\ \hline
        & \multicolumn{3}{|c||}{Helicity $\mathrm{W\rightarrow \ell \nu}$ }
        & \multicolumn{2}{c|} {Helicity W$\rightarrow$ hadrons}  \\ 
      \hline \hline     
 & $-1$ &$+1$& $0$ & $\pm 1$&$0$ \\
      \hline \hline
Data  & 0.568 $\pm$ 0.071 & 0.212 $\pm$ 0.046 & 0.220 $\pm$ 0.077
          & 0.716 $\pm$ 0.086 & 0.285 $\pm$ 0.084 \\  \hline
    
MC    & 0.56 & 0.18 & 0.26 & 0.74 & 0.26 \\  \hline \hline

        \multicolumn{6}{|c|}{$\sqrt{s}=183$ \GeV{}} \\ \hline \hline
Data  & 0.56 $\pm$ 0.14 & 0.10 $\pm$ 0.08 & 0.34 $\pm$ 0.15 
          & 0.75 $\pm$ 0.17 & 0.25 $\pm$ 0.17 \\  \hline
    
MC        & 0.53 & 0.20 & 0.27 & 0.73 & 0.27 \\  \hline

    \end{tabular}
    \icaption{Measured W helicity fractions
        for the leptonic and hadronic W decays for 
the $\sqrt{s}=189$ \GeV{} and $\sqrt{s}=183$ \GeV{} data samples.
The corresponding helicity fractions 
in the Standard Model as implemented in the KORALW Monte Carlo program 
where the statistical errors are negligible in comparison with the data
are also given.  
    \label{tab:table1}}
  \end{center}
\end{table}
\newpage
\begin{table}[htbp]
  \begin{center}
    \begin{tabular}{|c|c|c|c|}\hline
        \multicolumn{4}{|c|}{Fraction of longitudinally 
polarised W bosons}\\ \hline
      \hline \hline     
 &  W$\mathrm{\rightarrow \ell \nu}$   & W$\rightarrow$ hadrons & average \\
      \hline \hline
standard method                  & 0.220 $\pm$ 0.077 & 0.285 $\pm$ 0.084 & 
0.252 $\pm$ 0.057 \\ \hline
   \hline
background corrections           & 0.209--0.232  & 0.282--0.286 & 
0.241--0.258 \\
efficiency uncertainty         & 0.214        & 0.279       & 0.247 \\
calorimeter calibration (hadrons) & 0.197--0.215  & 0.282--0.300 
& 0.244--0.254 \\
integrated efficiency correction & 0.233        & 0.268       & 0.250 \\
analysis method                  & 0.237        & 0.279       & 0.258 \\
\hline
    \end{tabular}
    \icaption{Measurements of the fraction of longitudinally 
polarised W bosons for leptonic and hadronic W decays from the 
$\sqrt{s}=189$ \GeV{} data sample investigating 
various sources of systematics.
    \label{tab:table3}}
  \end{center}
\end{table}

\newpage
\begin{table}[htbp]
  \begin{center}
    \begin{tabular}{|c|c|c|c||c|c|}\hline
        \multicolumn{6}{|c|}{$\sqrt{s}=183+189$ \GeV{} data} \\ \hline
        & \multicolumn{3}{|c||}{Helicity $\mathrm{W\rightarrow \ell \nu}$ }
        & \multicolumn{2}{c|} {Helicity $\mathrm{W}\rightarrow$ hadrons }  \\ 
      \hline \hline     
$\cos \theta_{\mathrm{W}^{-}}$ & $-1$ &$+1$& $0$ & $\pm 1$&$0$ \\
      \hline \hline
$-1.0$ -- $-0.4$  & 0.27  $\pm$ 0.12  & 0.45   $\pm$ 0.22  & 0.28   $\pm$ 0.23 
& 0.87   $\pm$ 0.28  & 0.13   $\pm$ 0.28 \\   
$-0.4$ -- $\pho  0.3$  & 0.40 $\pm$ 0.09 & 0.23 $\pm$ 0.08 &  0.37  $\pm$ 0.12
& 0.94   $\pm$ 0.16  &  0.06 $\pm$ 0.15 \\   
$\pho  0.3$ -- $\pho  1.0$  & 0.66 $\pm$ 0.08 & 0.08 $\pm$ 0.04 & 
0.26 $\pm$ 0.08 & 0.75   $\pm$ 0.11  & 0.23 $\pm$ 0.10 \\   
\hline
\hline
        \multicolumn{6}{|c|}{$\sqrt{s}=183+189$ \GeV{} KORALW MC} \\ \hline
        & \multicolumn{3}{|c||}{Helicity $\mathrm{W\rightarrow \ell \nu}$ }
        & \multicolumn{2}{c|} {Helicity W$\rightarrow$ hadrons }  \\ 
      \hline \hline     
$\cos \theta_{\mathrm{W}^{-}}$ & $-1$ &$+1$& $0$ & $\pm 1$&$0$ \\
      \hline \hline
$-$1.0 -- $-$0.4  &  0.13  & 0.45   & 0.42 
&0.58  & 0.42   \\   
$-$0.4 --  \pho 0.3  & 0.42 & 0.29 & 0.29
& 0.71  & 0.29 \\   
\pho 0.3 -- \pho 1.0  & 0.67 & 0.10 & 0.23
& 0.77 & 0.23 \\   
\hline
    \end{tabular}
    \icaption{Same as Table 1, except in this case the helicity 
fractions are given as a function of 
$\cos \theta_{\mathrm{W}^{-}}$ and combining 
the $\sqrt{s}=183$ \GeV{} and $\sqrt{s}=189$ \GeV{}
data and Monte Carlo.
    \label{tab:table2}}
  \end{center}
\end{table}
\clearpage
%%%%%%%%%%%%%%%%%%%%%%%%%%%%%%%%%%%%%%%%%%%%%%%%%%%%%%%%%%%%%%%%%%%%%%%%%%%%%%
%\subsection*{Figures}

%%%%%%%%%%%%%%%%%%%%%%%%%%%%%%%%%%%%%%%%%%%%%%%%%%%%%%%%%%%%%%%%%%%%%%%%%%%%%%%

\newpage

\newpage

\newpage
\newpage

%%%%%%%%%%%%%%%%%%%%%%%%%%%%%%%%%%%%%%%%%%%%%%%%%%%%%%%%%%%%%%%%%%%%%%%%%%%%%%%

%
%%%%%%%%%%%%%%%%%%%%%%%%%%%%%%%%%%%%%%%%%%%%%%%%%%%%%%%%%%%%%%%%%%%%%%%%%%%%%%
% Author List
%%%%%%%%%%%%%%%%%%%%%%%%%%%%%%%%%%%%%%%%%%%%%%%%%%%%%%%%%%%%%%%%%%%%%%%%%%%%%%
%

\end{document}